\documentclass[12pt]{article}
\usepackage{epsfig}
\usepackage[english]{babel}
\usepackage[T1]{fontenc}
\usepackage{graphicx}
\usepackage{amsmath, amssymb, latexsym, amsopn}
\usepackage{mathtools}
\usepackage{authblk}
\usepackage{hyperref}

\newcommand{\av}[1]{\langle{#1}\rangle}

\title{Gravitational entropy and the cosmological no-hair conjecture}

\author[1]{Krzysztof Bolejko\thanks{krzysztof.bolejko@sydney.edu.au}}
\affil[1]{Sydney Institute for Astronomy, School of Physics, A28, The University of Sydney, NSW, 2006, Australia
}

\begin{document}

\maketitle

\begin{abstract}
The gravitational entropy and no-hair conjectures seems to 
predict contradictory future states of our Universe.
The growth of the gravitational entropy is associated with the growth of inhomogeneity, while the no-hair conjecture argues that a universe dominated by dark energy should asymptotically approach a homogeneous and isotropic de Sitter state. The aim of this paper is to study these two conjectures. The investigation 
is based on the Simsilun simulation, which simulates the universe using the approximation of the Silent Universe. The Silent Universe is a solution to the Einstein equations that assumes irrotational, non-viscous, and insulated dust, with vanishing magnetic part of the Weyl curvature. 
The initial conditions for the Simsilun simulation are sourced from the Millennium simulation, which results with a realistically appearing but relativistic at origin simulation of a universe. The Simsilun simulation is evolved from the early universe ($t = 25$ Myr) till far future ($t = 1000$ Gyr). The results of this investigation show that both conjectures are correct.
On global scales, a universe with a positive cosmological constant and non-positive spatial curvature does indeed approach the de Sitter state. At the same time it keeps generating the gravitational entropy.
\end{abstract}

\section{Introduction}\label{intro}

Gravitational systems with their long range gravitational interactions
have different properties than thermodynamic systems that we typically encounter on Earth.
For example, a typical sequence of events for a gas injected into an empty a box 
is to evolve from clumpiness towards homogeneity.
However, for a system that is dominated by gravity, a reverse sequence of events
is typically observed, and so the system evolves from homogeneity towards clumpiness \cite{1989NYASA.571..249P}.
It is still debatable, whether one can 
define a gravitational entropy, i.e. a quantity that is analogous to the thermodynamic entropy, which would encapsulate a typical behaviour of all gravitational systems \cite{2013CQGra..30l5009C}. 
The issue of gravitational entropy is inevitably related to the issue of the cosmological arrow of time. 

When in 1980s, Penrose  postulated the {\em Weyl Curvature Hypothesis} \cite{1989NYASA.571..249P,1989esnm.book.....P},
the debate started whether this could serve as a meaningful measure of the cosmological arrow of time. 
The Weyl Curvature Hypothesis states that the universe starts with 
zero Weyl curvature and evolves to a state dominated by the Weyl curvature.
The magnitude of the Weyl Curvature could thus be related to the arrow of time.
However, the first attempt to link the Weyl Curvature with the gravitational entropy were not fully successful,
as such a definition of the gravitational entropy has problems with decaying modes \cite{1997PhRvD..55.1948R}. Similarly, the definition of the gravitational entropy based on the ratio of the Weyl to Ricci curvatures has problems with radiation \cite{1987PhLA..122..305B}.

However, when Senovilla showed that the Bel-Robinson tensor can be used to construct a reasonable  measure of the 
 `energy' of the gravitation field \cite{2000CQGra..17.2799S}, this promoted attempts to define the
gravitational entropy based on the Bel-Robinson tensor \cite{2000PhRvD..62d4009P,2006IJTP...45.1258P}. Recently, Clifton, Ellis, and Tavakol showed how using the Bel-Robinson tensor one can construct an `effective energy-momentum tensor' of the gravitational field \cite{2013CQGra..30l5009C}.
This allowed them to derived the formula for the gravitational entropy. The derivation was obtained in a similar way to a derivation of the  thermodynamic entropy based on the  energy-momentum tensor \cite{2013CQGra..30l5009C}.
This new formula seems to meet requirements for the gravitational entropy, such as: (i) suitable limit for the 
Bekenstein-Hawking entropy of black holes \cite{1973PhRvD...7.2333B,1973CMaPh..31..161B}, and (ii) for cosmological systems the increase of the gravitational entropy 
is associated with the growth of cosmic structures
 \cite{2013CQGra..30l5009C,2013PhRvD..88f3529B,2014CQGra..31g5021S,2014AN....335..587S,2015CQGra..32p5012S,2015GReGr..47..114M}.

The procedure of defining the gravitational entropy based on the Bel-Robinson tensor requires a procedure of fining a {\em square root} of the Bel-Robinson tensor, which can only be done within spacetime of Petrov type D and N \cite{2000CQGra..17.2799S}, and so this limits the applicability of such a procedure.
However, this is not the only approach to the gravitational entropy that is being considered in the literature.
Another approach, which seems quite promising for cosmological systems is the one based on the 
Kullback-Leibler relative information entropy. Such a procedure has been  suggested by Hosoya, Buchert, and Morita  \cite{2004PhRvL..92n1302H}.
The gravitational entropy which is defined in this way is conjectured to grow in generic situations due to negative feedback of open gravitational systems, which is proved
to hold for linear perturbations of an Einstein-de Sitter background model, and exact
Lema\^itre--Tolman models \cite{2010AIPC.1241.1074M}.
It has also been show that in the cosmological context the 
Kullback-Leibler relative information entropy can be well approximated by the
R\'enyi relative entropy \cite{2016PhLB..758....9C},
and that it can be linked to Weyl curvature  \cite{2012PhRvD..86h3539L}.
To distinguish this approach from the other, let us denote the entropy defined based on the information entropy as the {\em HBM gravitational entropy} and
the gravitational entropy that is derived from the Bel-Robinson tensor as the {\em CET gravitational entropy}.

In  parallel to the gravitational entropy and Weyl Curvature Hypothesis,
the idea of cosmic {\em inflation} was being developed  \cite{1979JETPL..30..682S,1981PhRvD..23..347G,1982PhLB..108..389L}.
During the cosmic inflation the dynamics of the universe is dominated by the scalar filed 
and the universe rapidly evolves towards the de Sitter state.
This observation, together with number of other studies \cite{PhysRevD.28.2118,1984PhLA..103..315B,1991AnP...503..518P}, formed foundations for the {\em cosmic no-hair conjecture}.
The cosmic no-hair conjecture states that a universe with a positive cosmological constant evolves towards the de Sitter state.
Although, for some configurations this does not occur \cite{1986PhLB..180..335B,GOTZ1988129}, in general it is expected that our Universe will eventually evolve towards the de Sitter state.
This is in contrast with what one expects based on the requirement of the growth of the gravitational entropy. 
This apparent contradiction motivates the research of this paper. The structure of this paper is as follows:
Sec. \ref{silent} describes the Silent Universe;
Sec. \ref{nohair} sketches a derivation of the cosmological no-hair conjecture;
Sec. \ref{gravent} derives the formula for the gravitational entropy of the silent universe;
Sec. \ref{simsilun} presents and applies the Simsilun simulation to investigate the production rate of the gravitational entropy and future properties of the universe that is dominated by the cosmological constant;
Sec. \ref{conclusions} concludes the results.

\section{Silent Universe}\label{silent}

\subsection{Relativistic evolution of irrotational and insulated cosmic dust}

We first assume that the gravitational field is 
sourced by  the irrotational (no vorticity) and insulated (no heat transfer) dust with a cosmological constant.
We then thread the spacetime with lines
that are tangent to the flow of matter $u^a$, and slice the spacetime with surfaces that are orthogonal
to $u^a$. This results with 1+3 split and comoving coordinates \cite{1971grc..conf..104E,2009GReGr..41..581E}. 
Applying the energy-momentum conservation equations $T^{ab}{}_{;b} =0$,
the Ricci identities $u_{a;d;c}-u_{a;c;d} = R_{abcd} u^b$, and the Bianchi identities $R_{ab[cd;e]} = 0$, the evolution of the system is given by

 \begin{align}
& \dot{\rho}  + \Theta \, \rho  = 0, \label{ffe1} \\   
& \dot{\Theta}  = - \frac{1}{3}\,\Theta^2-  \frac{1}{2}\,\rho-
2\sigma^2+ \Lambda,  \\
&  \dot{\sigma}_{\langle ab\rangle}  = - \frac{2}{3}\,\Theta\sigma_{ab}-
\sigma_{c\langle a}\sigma^c{}_{b\rangle} - E_{ab},\\
&  \dot{E}_{\langle ab\rangle}  = -\Theta E_{ab}-
 \frac{1}{2}\,\rho \, \sigma_{ab}+  {\rm curl} \, H_{ab}  + 3\sigma_{\langle
a}{}^c \, E_{b\rangle c},   \\
&  \dot{H}_{\langle ab\rangle} = -\Theta H_{ab}- {\rm curl} \, E_{ab} +3\sigma_{\langle a}{}^cH_{b\rangle c}. \label{ffe10} 
\end{align}
In addition there are spatial constraints that follow from the spatial parts of the Ricci and Bianchi identities

\begin{align}
& D^b \sigma_{ab} = \frac{2}{3} D_a \Theta, \label{brc1} \\
& H_{ab} = {\rm curl} \, \sigma_{ab}, \label{brc3} \\
& D^b E_{ab} = \frac{1}{3} D_a \rho  + \epsilon_{abc} \sigma^b{}_{d} \, H^{cd},  \label{brc4} \\
& D^b H_{ab} =  - \epsilon_{abc} \sigma^b{}_{d} E^{cd}.  \label{brc5}
\end{align}

The above set of equations is equivalent to the Einstein equations
$G_{ab} - \Lambda g_{ab} = T_{ab}$. However, instead of solving the 
Einstein equations directly, we deal with equations that describe  properties of dust (i.e.~$\rho$), its velocity field (i.e.~$\Theta$, $\Sigma$),
and spacetime geometry (i.e.~$E_{ab}$ and $H_{ab}$).
Note that the constant $\kappa = 8 \pi G/c^4$ is assumed to be 1, which for a pressureless and  insulated dust
is equivalent to rescaling of density, i.e. $\kappa \rho \to \rho$.

\subsection{Silent universe}

The above equations describe a general relativistic evolution of irrotational and insulated cosmic dust.
In the absence of pressure gradients, there are no sound waves making this universe almost `silent'. In order to enforce strict `silence' and prevent any communication between the worldliness, we need to put $H_{ab} = 0$, which will prevent propagation of gravitational waves.
In such a case the above system of equations is expected to describe spacetimes that are Petrov D  \cite{1997CQGra..14.1151V} with the shear and electric part of the Weyl tensors taking the form

\begin{equation}
\sigma_{ab} = \Sigma \, {\rm e}_{ab}, \quad E_{ab} = {\cal W} \, {\rm e}_{ab}, \label{brcs}
\end{equation}
where ${\rm e}_{ab} = h_{ab} - 3 z_a z_b$ where $z^a$ is a space-like unit vector aligned with the Weyl principal tetrad. As a result the fluid equations (\ref{ffe1})--(\ref{ffe10})
reduce only to 4 scalar equations \cite{1995ApJ...445..958B,1997CQGra..14.1151V}

\begin{align}
 & \dot \rho = -\rho\,\Theta, \label{rhot}\\
& \dot \Theta = -\frac{1}{3}\Theta^2-\frac{1}{2}\,\rho-6\,\Sigma^2 + \Lambda,\label{thtt}\\
& \dot \Sigma = -\frac{2}{3}\Theta\,\Sigma+\Sigma^2-{\cal W},\label{sigmat}\\
& \dot{ {\cal W}} = -\Theta\, {\cal W} -\frac{1}{2}\rho\,\Sigma-3\Sigma\,{\cal W},\label{weyt}
\end{align}
with the spatial constraints
\begin{align}
& D^b \sigma_{ab} = \frac{2}{3} D_a \Theta, \label{sigmac} \\
& D^b E_{ab} = \frac{1}{3} D_a \rho. \label{weyc}
\end{align}

\section{Cosmological no-hair conjecture}\label{nohair}

This section presents a heuristics derivation of the cosmological no-hair conjecture. 
This derivation should not be treated as a mathematically complete derivation, rather it should be treated as a  point of reference for a further discussion. For a more strict derivation, the Reader is refereed to Refs. \cite{GOTZ1988129,1991AnP...503..518P,HadzicSpeck,Oliynyk2016}.

Assuming a non-positive spatial curvature
\begin{equation}
{\cal R} \leq 0,
\label{coshair}
\end{equation}
it follows from eq. (\ref{thtt}) that 
\begin{equation} \dot \Theta \leq -\frac{1}{3}\Theta^2 + \Lambda \leq 0.
\label{decexp}
\end{equation}
The first inequality follows from the fact that $\rho \geq 0$ and $\Sigma^2 \geq 0 $,
and the second follows from the Hamiltonian constraint
\begin{equation}
- \frac{1}{3} \Theta^2 +  \Lambda =  - \rho  - 3 \, \Sigma^2 + \frac{1}{2} {\cal R},
\label{hamcon}
\end{equation}
which shows that for ${\cal R} \leq 0$ the left hand side of the above equation cannot be positive.
Thus, from eq. (\ref{decexp}) it follows that the expansion rate decreases and 
\begin{equation}
 \Theta \to \sqrt{ 3 \Lambda}. 
\label{deSlimit}
\end{equation}
If this happens then, as follows from the Hamiltonian constraint (\ref{hamcon})
\[   \rho + 3 \Sigma^2  \to \frac{1}{2} {\cal R}. \]
However, since ${\cal R} \leq  0$, this implies that
\[ \rho \to 0,  \quad \Sigma^2 \to 0, \quad {\cal R} \to 0,  \]
and from (\ref{weyt}) 
\[ {\cal W} \to 0. \]
As a result, the universe asymptotically approaches the de Sitter space --- spatially flat, homogeneous and isotropic FLRW model, with the expansion rate $\Theta =\sqrt{ 3 \Lambda}$.

\vspace{0.3cm}
\noindent
\underline{Conjecture 1} (cosmological no-hair conjecture)
\vspace{0.1cm}

\noindent
{\em A universe with a non-positive spatial curvature and positive cosmological constant asymptotically evolves towards the de Sitter universe.}
\vspace{0.2cm}

\section{Gravitational entropy}\label{gravent}

\subsection{CET gravitational entropy}

In analogy to thermodynamic and relativistic systems, 
one can define an `effective' energy-momentum tensor of the free gravitational field \cite{2013CQGra..30l5009C}
\begin{equation}
 {\cal T}^{ab} = \rho_{{\rm grav}} u^a u^b + p_{{\rm grav}} h^{ab} + \Pi_{{\rm grav}}^{ab} +  2 q_{{\rm grav}}^{(a} u^{b)}, 
\end{equation}
which for Petrov D spacetimes is \cite{2013CQGra..30l5009C}
\begin{align}
& q_{{\rm grav}}^{a} = 0, \nonumber \\
& p_{{\rm grav}} = 0, \nonumber \\
& \Pi_{{\rm grav}}^{ab} = \frac{\alpha}{4\pi} | \Psi_2 | (x_a x_b + y_a y_b - z_a z_b + u^a u^b) , \nonumber \\
& \rho_{{\rm grav}} = \frac{\alpha}{4\pi} | \Psi_2 | = \frac{\alpha}{4\pi}  {\cal W} \; {\rm sgn}( {\cal W}), \nonumber \\
& T_{{\rm grav}} = \frac{1}{2\pi} \left( \frac{1}{3} \Theta - 2 \Sigma \right),
\end{align}
where $\Psi_2$ is the conformal Newman-Penrose invariant, $\alpha$ is a constant, and ${\rm sgn}( {\cal W})$ is the sign of ${\cal W}$, i.e. 
${\rm sgn}( {\cal W}) = |{\cal W}|/{\cal W}$.

 The growth of the gravitational entropy is thus

\begin{equation}
\dot{s}_{\text{\tiny CET}} = \frac{1}{T_{{\rm grav}}} \big( \rho_{{\rm grav}} \, \delta v \big)\dot{},
\end{equation}
where $\delta v$ is the local volume element. Since the  rate of change of volume is proportional to the expansion rate
\begin{equation}
\delta \dot{ v} = \delta v \, \Theta,
\end{equation}
thus
\begin{equation}
\dot{s}_{\text{\tiny CET}} = \frac{\delta v}{T_{{\rm grav}}} \big( \dot{\rho}_{{\rm grav}} + {\rho}_{{\rm grav}} \Theta \big).
\end{equation}
Finally using eq. (\ref{weyt}) 
\begin{equation}
\dot{s}_{\text{\tiny CET}} = -  \alpha  \frac{3}{4} \frac{ \rho \Sigma + 6 \Sigma {\cal W}}{  | \Theta - 6 \Sigma | } \; {\rm sgn}( {\cal W}) \, \delta v.
\end{equation}
Below, out of convenience, the arbitrary constant $\alpha$ is set to
\[ \alpha = \frac{4}{3 H_0^2}, \]
\noindent where $H_0$ is the Hubble constant ($H_0$ has the same units as $\Theta$ and $\Sigma$ and $\rho^{1/2}$).  Thus, for the silent universe, the growth of the gravitational entropy is
\begin{equation}
\dot{s}_{\text{\tiny CET}} = -   \frac{\Sigma}{H_0^2} \frac{ \rho + 6 {\cal W}}{  | \Theta - 6 \Sigma | } \frac{ {\cal W} }{ | {\cal W}|} \, \delta v. \label{silgravent}
\end{equation}
Integrating over the whole domain ${\cal D}$, the change of rate
of the gravitational entropy of the silent universe is

\begin{equation}
\dot{S}_{\text{\tiny CET}} = - \int_{\cal D} \delta v \,\frac{\Sigma}{H_0^2} \frac{ \rho + 6 {\cal W}}{  | \Theta - 6 \Sigma | } \frac{ {\cal W} }{ | {\cal W}|}. \label{CETintgravent}
\end{equation}

\subsection{HBM gravitational entropy}

In analogy to information entropy when
the relative entropy measures how one distribution diverges from the other,
Hosoya, Buchert, and Morita   suggested 
to define the gravitational entropy as a measure of divergence of the matter density field from its global average \cite{2004PhRvL..92n1302H}

\begin{equation}
S_{\text{\tiny HBM}} =\int_{\cal D} \delta v \, 
 \rho \ln \frac{\rho}{\av{\rho}_{\cal D}}, \label{hbmint}
\end{equation}
where $\av{\rho}_{\cal D}$ is the volume average density  
\begin{equation}
\av{\rho}_{\cal D} = \frac{1}{V_{\cal D}} \int_{\cal D} \delta v \, \rho.
\end{equation}
To make the units of $S_{\text{\tiny HBM}}$ the same as of $S_{\text{\tiny CET}}$
we scale it by $H_0^3$ and so the rate of change of the HBM gravitational entropy
can be written as  \cite{2004PhRvL..92n1302H}
\begin{equation}
\dot{S}_{\text{\tiny HBM}} = 
- \frac{1}{H_0^3} \left( \int_{\cal D} \delta v \, \rho \Theta \right) 
+ \frac{1}{H_0^3 V_{\cal D}} \left( \int_{\cal D} \delta v \, \rho \right)
\left( \int_{\cal D} \delta v \, \Theta \right).
\label{HBMintgravent}
\end{equation}

\subsection{Gravitational entropy in the early universe, i.e. 
small perturbations around the Einstein-de Sitter model}\label{edse}

The early universe is often described using the  Einstein-de Sitter model.
The reason for that is that the contribution for spatial curvature 
${\cal R}$ and the cosmological constant $\Lambda$ is negligible small compared
to the contribution from matter energy density $\rho$. 
In addition, if the distribution of matter is sufficiently uniform
(standard assumption in cosmology) then it seems that the application of the Einstein-de Sitter model 
to describe the properties of the early universe is justified. 
In such a case, the Hamiltonian constraint (\ref{hamcon}) reduces to
\begin{equation}
3\bar{\rho} = \bar{\Theta}^2, 
\end{equation}
where the bar is used to denote the Einstein-de Sitter model. 
The early universe is not strictly spatially homogeneous and isotropic, but there are perturbations
around the  Einstein-de Sitter background

\[ \rho = \bar{\rho}  + \Delta \rho \quad {\rm ~and} \quad \Theta = \bar{\Theta} +  \Delta \Theta. \]
If the perturbations are small and dominated by the growing mode then \cite{1980lssu.book.....P} 
\[  \Delta \rho = \bar{\rho} \, \delta \quad {\rm ~and} \quad \Delta \Theta = - \frac{1}{3}  \bar{\Theta} \, \delta. \]
Inserting (\ref{brcs}) to  (\ref{sigmac}) and (\ref{weyc})
\begin{align}
& {\rm e}_{ab} \, D^b \Sigma + \Sigma \, D^b {\rm e}_{ab} = \frac{2}{3} D_a \Theta  = - \frac{2}{9} \bar{\Theta}^2 D_a \delta, \\
& {\rm e}_{ab} \, D^b {\cal W} + {\cal W} D^b {\rm e}_{ab} = \frac{1}{3} D_a \rho =  
\frac{1}{3} \bar{\rho} D_a \delta_i  = \frac{1}{9} \bar{\Theta}^2   D_a \delta.
\end{align}
Comparing the right hand sides of the above equations and neglecting higher order terms, such as 
$\Sigma \, \delta$ we arrive at
\begin{equation} 
{\cal W} =  - \frac{1}{2} \bar{\Theta} \, \Sigma =  - \frac{3}{2}  \frac{\Sigma}{\bar{\Theta}} \, \bar{\rho}.
\label{wsig}
\end{equation}
Inserting above to (\ref{CETintgravent})
\begin{equation}
\dot{S}_{\text{\tiny CET}} =  \int_{\cal D} \delta v \, \frac{3}{2} \frac{\Sigma^2}{H_0^2}  \frac{ \bar{\rho}+ 6 {\cal W}}{ | \bar{\Theta} - 6 \Sigma | }
\frac{ \bar{\rho} }{ | {\cal W}| \, \bar{\Theta}} 
\approx \int_{\cal D} \delta v \, \frac{3}{2} \frac{\Sigma^2}{H_0^2}  \frac{ \bar{\rho}^2}{  \bar{\Theta}^2  | {\cal W}| },
\label{cetlin}
\end{equation}
where the higher order terms have been dropped.
The above formula, despite appearance of a second order quantity (i.e. $\Sigma^2$), is first-order in perturbations (${\cal W} \sim \Sigma$), however, unlike a first-order quantity it does not vanish after averaging over the whole domain, as the integrand is positive. 

In the case of the HBM gravitational entropy, for small and compensated perturbations, the integral (\ref{HBMintgravent}) reduces to 
\begin{equation}
\dot{S}_{\text{\tiny HBM}} = - \frac{1}{H_0^3}
\int_{\cal D} \delta v \, \Delta \rho \, \Delta \Theta \approx \frac{1}{9} \, \frac{\bar{\Theta}^3}{H_0^3} \int_{\cal D} \delta v \, \delta^2.
\end{equation}
Unlike the CET, this is truly the second-order quantity (the first order quantities has been integrated out) and 
within the applicability of the above assumptions, the growth rate of the HBM gravitational entropy is positive.

\subsection{Gravitational entropy conjecture}

In both cases (CET and HBM), the growth of the gravitational entropy vanishes in the FLRW case. In the FLRW case the shear $\Sigma$ and Weyl curvature ${\cal W}$ vanish
and so the integrand (\ref{CETintgravent}) vanishes leading to 
 $\dot{S}_{\text{\tiny CET}} = 0$. 
 For the HBM case, in the FLRW regime, the first term in (\ref{HBMintgravent}) is equal to the second one and so  $\dot{S}_{\text{\tiny HBM}} = 0$.
Thus, as expected: {\em the FLRW models do not produce the gravitational entropy}.
Treating this as a logical proposition, the negation of the reverse is also a logically correct statement, hence:

\vspace{0.3cm}
\noindent
\underline{Proposition 1}
\vspace{0.1cm}

\noindent
{\em Any universe that generates gravitational entropy cannot belong to a family of spatially homogeneous and isotropic FLRW models.}
\vspace{0.2cm}

As shown in Sec.~\ref{edse}, for small perturbations around the Einstein--de Sitter model the production rate of the gravitational entropy is positive.
Thus, it seems that it is reasonable to expect that a realistic model of a universe 
can be characterised with  a positive rate of change of the gravitational entropy. Therefore, the following conjecture is postulated:

\vspace{0.3cm}
\noindent
\underline{Conjecture 2} (cosmological gravitational entropy conjecture)
\vspace{0.1cm}

\noindent
{\em The evolution of the universe proceeds in such a way that it keeps generating the gravitational entropy.}
\vspace{0.2cm}

\begin{figure}[h!]
\begin{center}
\includegraphics[scale=0.9]{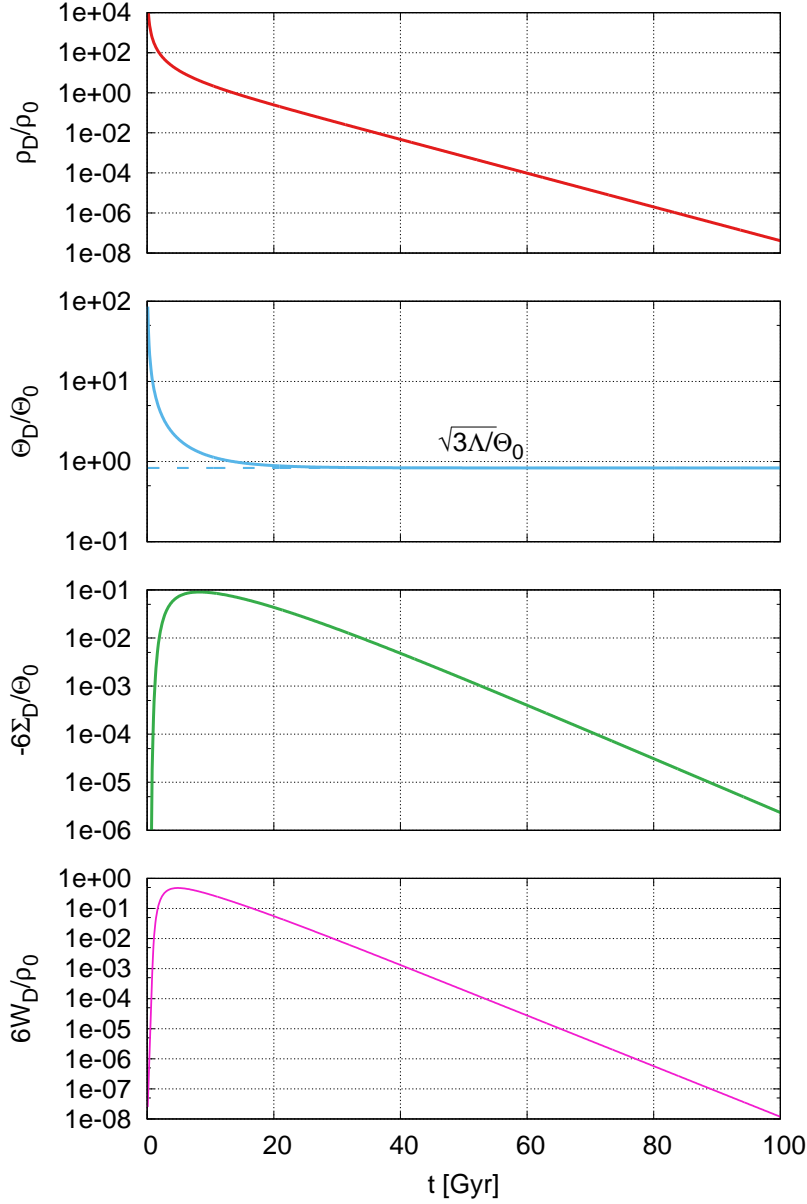}
\end{center}
\caption{The evolution of the volume averages of: the density field normalised by the present-day density ({\em upper most panel}); expansion rate normalised by the present-day expansion rate ({\em second upper panel}); shear normalised by the one sixth of the present-day expansion rate (cf. (\ref{silgravent}))
and multiplied by $-1$ so that is can be presented in the log-y plot ({\em second lower panel});
the Weyl curvature normalised by the one sixth of the present-day density ({\em lower most panel}).
As seen, asymptotically the system approaches the de Sitter state, i.e. $\rho_{\cal D} \to 0$, $\Sigma_{\cal D} \to 0$, ${\cal W}_{\cal D} \to 0$, and $\Theta \to \sqrt{3 \Lambda}$.
Also, the product of the shear and Weyl curvature is negative $\Sigma {\cal W} < 0$
which as follows from (\ref{silgravent}) should imply a non-negative rate of change of the gravitational entropy.} 
\label{fig1}
\end{figure}

\begin{figure}
\begin{center}
\includegraphics[scale=0.95]{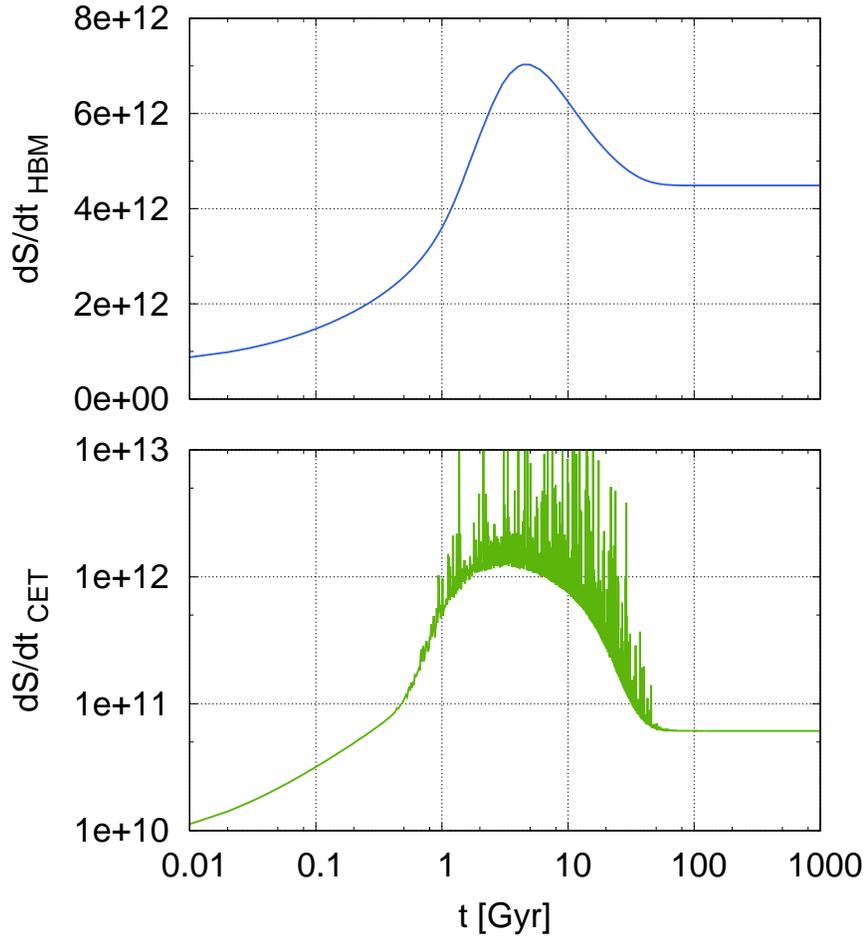}
\end{center}
\caption{The rate of change of the gravitational entropy within the whole domain of the Simsilun simulation. {\em Upper Panel}: change of rate of the HBM gravitational entropy;
{\em Lower Panel}: change of rate of the CET gravitational entropy.}
\label{fig2}
\end{figure}

\clearpage

The above is just a conjecture as it is based on properties of small perturbations.
In the next section we will test this conjecture by performing simulation that will allow us to trace the evolution far into the non-linear regime.

\section{Results}\label{simsilun} 

Conjecture 2 (the cosmological gravitational entropy conjecture) together with Proposition 1
seem to be in contradiction with Conjecture 1 (the cosmological no-hair conjecture), which postulates that a universe 
with a positive cosmological constant will end up as a homogeneous and isotropic de Sitter model. 
In this Section we will test these conjectures using the Simsilun simulation  \cite{simsulun.paper}.

\subsection{The Simsilun simulation}

The Simsilun simulation is based on the code \textsl{simsilun}\footnote{\url{https://bitbucket.org/bolejko/simsilun}}.
The description of the code, equations, and its  applications are described in the `Methods Paper' \cite{simsulun.paper}.
The Methods Paper describes how one can use the 
 Millennium simulation \cite{2005Natur.435..629S,2009MNRAS.398.1150B,2013MNRAS.428.1351G}
to set up the initial conditions for the code \textsl{simsilun}.
The Simsilun simulation is based on solving eqs. (\ref{rhot})--(\ref{weyt}), with the initial conditions given by 

\begin{align}
& \rho_i = \bar{\rho} + \Delta \rho = \bar{\rho} \, ( 1 + \delta_i), \label{rhoi} \\
& \Theta_i = \bar{\Theta} + \Delta \Theta =  \bar{\Theta} \, ( 1 - \frac{1}{3}  \delta_i), \label{thti} \\
& \Sigma_i = - \frac{1}{3} \, \Delta \Theta = \frac{1}{9}  \bar{\Theta} \, \delta_i, \label{sigmai} \\
& {\cal W}_i = - \frac{1}{6} \bar{\rho} \, \delta_i, \label{weyi}
\end{align}
where the subscript $i$ denotes the initial values, and $\delta_i$ is the initial density contrast
sourced from the Millennium Simulation \cite{2005Natur.435..629S,2009MNRAS.398.1150B,2013MNRAS.428.1351G}. Here we use the MField, which stores the matter distribution
smoothed with a Gaussian kernel of radius $2.5\,h^{-1}$ Mpc.
Since the MField consists of $256^3$ cells thus the resulted simulation, referred to as the Simsilun simulation, consists of 16,777,216 worldlines. 
In addition, the virialisation mechanism no 1 is implemented, whose technical details are is described in Sec. 3 in the Methods Paper \cite{simsulun.paper}.
The  Millennium Simulation is based on the $\Lambda$CDM model with
$\Omega_M = 0.25$, $\Omega_\Lambda = 0.75$, and $H_0 = 73.0$ km s$^{-1}$ Mpc$^{-1}$.
This background model meets all the requirements for the applicably of the cosmic no-hair conjecture:
it contains a positive cosmological constant and non-positive spatial curvature \cite{1991AnP...503..518P},
and it should asymptotically approaches the de Sitter solution.

\subsection{Gravitational entropy and the cosmological ``no-hair'' conjectures}

We calculate the evolution of the universe as described in the Methods Paper \cite{simsulun.paper}, but instead of stopping at the present day, the evolution of the system is followed till $t = 1000$ Gyr. Also, in addition to the evolution eqs. (\ref{rhot})--(\ref{weyt}), the volume of each element (cell) $\delta v$ is evolved using 
\begin{equation}
\delta \dot{v} = \delta v \, \Theta.
\end{equation}
Finally, the average properties of the Simsilun simulation are evaluated using the volume averages

\begin{align}
& \rho_{\cal D} = \av{\rho}_{\cal D} = \frac{ \sum_j \, \delta v_j \, \rho_j}{ \sum_j \, \delta v_j}, \\
& \Theta_{\cal D} = \av{\Theta}_{\cal D} = \frac{ \sum_j \, \delta v_j \, \Theta_j}{ \sum_j \, \delta v_j}, \\
& \Sigma_{\cal D} = \av{\Sigma}_{\cal D} = \frac{ \sum_j \, \delta v_j \, \Sigma_j}{ \sum_j \, \delta v_j}, \\
& {\cal W}_{\cal D} = \av{{\cal W}}_{\cal D} = \frac{ \sum_j \, \delta v_j \, {\cal W}_j}{ \sum_j \, \delta v_j},
\end{align}
where $\rho_j$, $\Theta_j$, $\Sigma_j$, and ${\cal W}_j$ are quantities evaluated at each
cell, whose volume is $\delta v_j$. The volume of the domain of averaging ${\cal D}$ is the total volume of the Simsilun simulation and is evaluated as
\begin{equation}
V_{\cal D}  = \sum_j \, \delta v_j.
\end{equation}

The volume averaged properties of the Simsilun simulation and their evolution is presented in Fig. \ref{fig1}. The evolution has been evaluated till $t=1000$ Gyr (for clarity of presentation, Fig. \ref{fig1} presents only evolution till $t=100$ Gyr). Also, the evolution of shear $\Sigma$ is multiplied by $-1$ so that is can be presented in the log-y plot. 
The presented results show that the volume averaged properties of the Simsilun simulation asymptotically approach the de Sitter state, i.e. $\rho_{\cal D} \to 0$, $\Sigma_{\cal D} \to 0$, ${\cal W}_{\cal D} \to 0$,
and $\Theta \to \sqrt{3 \Lambda}$. Thus, these results seem to confirm the no-hair conjecture.

\begin{figure}
\begin{center}
\includegraphics[scale=0.95]{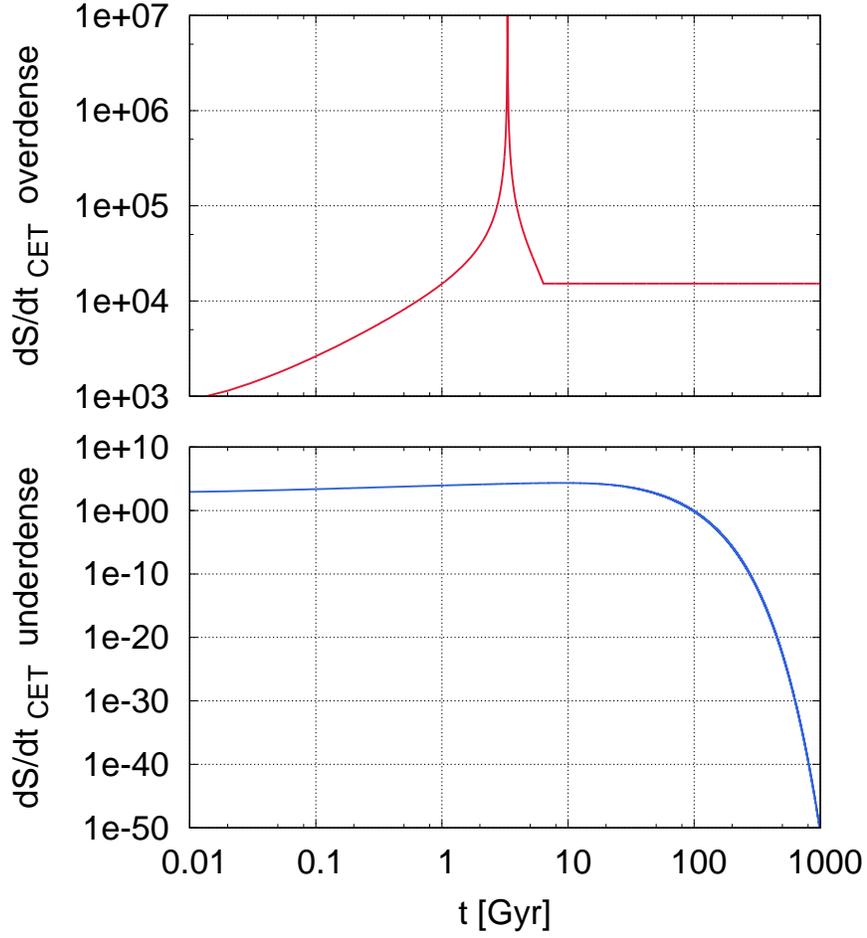}
\end{center}
\caption{The rate of change of the CET gravitational entropy for a single overdense cell ({\em Upper Panel}), and a single underdense cell ({\em Lower Panel}).}
\label{fig3}
\end{figure}

However, the results presented in Fig. \ref{fig2} also confirm the gravitational entropy conjecture as they show the positive rate of change of both CET and HBM gravitational entropies.
 The rate of change of the gravitational entropy
peaks a few billion years after the big bang, at the similar time scale 
when the shear and Weyl curvature reach their maximum amplitude (cf. Fig. \ref{fig1}). 
Thus, both formulae for the gravitational entropy (CET and HBM) provide a similar picture. This is not surprising as it has been shown that these two formulae
are correlated  \cite{2012PhRvD..86h3539L}. However, what is surprising is that when the universe approaches the de Sitter stage, the rate of change decreases, but does not vanish and after approximately 100 Gyr remains constant.

The CET gravitational entropy features a number of spikes. The origin of these spikes
is explained in Fig. \ref{fig3}, which presents the rate of change of the CET gravitational entropy for a single underdense and a single overdense cell. The initial conditions for the 
underdense and overdense regions are $\delta_i = -0.02$ and $\delta_i = 0.02$ respectively (cf. (\ref{rhoi})--(\ref{weyi})) and  the rate of change of their gravitational entropies  follows from (\ref{silgravent}).

For the overdense region (upper panel in Fig. \ref{fig3}),
approximately after 3 Gyr  of evolution the expansion rate slows down and $\Theta - 6 \Sigma \to 0$. This results with a spike (due to a finite numerical step of integration, the spike does not reach $\infty$, however in reality it does).
After the overdensity becomes virialised, the production rate of its gravitational entropy becomes constant. 
This is in contrast with the no-hair conjecture, but as stated in Sec.~\ref{nohair},
the cosmological no-hair conjecture does not apply if the 
spatial curvature is positive. For the region with a positive spatial curvature,
 which undergoes collapse and eventually virialisation, the future asymptotic state is not the de Sitter state and the production rate of the gravitational entropy does not asymptomatically vanish. 
The rate of change of the gravitational entropy does asymptotically vanish
for the underdense region (lower panel in Fig. \ref{fig3}), where 
the spatial curvature is negative and where the cosmological no-hair conjecture does apply,
which is also in agreement with results presented in Ref. \cite{2015CQGra..32p5012S} where the evolution of cosmic voids and their gravitational entropy was investigated using the Lema\^itre--Tolman model.

The results of Fig.~\ref{fig3} allow to understand the results of Fig.~\ref{fig2},
which show that the rate of change of the gravitational entropy does not asymptotically vanish
even though (as seen from Fig.~\ref{fig1}) the Simsilun simulation asymptotically approaches the de Sitter state.
The reason for this is following: the virialised overdense regions occupy little volume, and since they do not expand, thus with time, their contribution to the total volume is negligibly small, and therefore their contribution to the volume averages (cf. Fig. \ref{fig1}) is negligibly small. 
The Simsilun simulation consists of overdense and underdense regions. 
As a result, the volume of the Simsilun simulation is asymptotically dominated by underdense regions, but the production of the gravitational entropy is asymptotically dominated by overdense regions.
The reason why overdense regions produce the gravitational entropy is linked
to the fact that in the expanding universe virialised regions have non-zero shear (which is the source of the CET gravitational entropy, eq. (\ref{CETintgravent}))
and their density does not asymptotically approach the de Sitter limit (i.e. $\rho/\av{\rho}_{\cal D} \ne 1$ and asymptotically diverges, 
which sources the HBM entropy, eq. (\ref{hbmint})).
Therefore, even though the volume averaged properties of the  Simsilun simulation asymptotically approach the de Sitter state,  the rate of change of the gravitational entropy does not asymptotically vanish.

\section{Conclusions}\label{conclusions}

This paper investigated the cosmological no-hair and gravitational entropy conjectures. 
The investigation was based on the Simsilun simulation \cite{simsulun.paper}.
The Simsilun simulation simulates the universe using the approximation to the Einstein equations, which is based on the 
silent universes \cite{1995ApJ...445..958B,1997CQGra..14.1151V}.
In addition, the Simsilun simulation uses the initial data sourced from the Millennium simulation  \cite{2005Natur.435..629S,2009MNRAS.398.1150B,2013MNRAS.428.1351G}.

The obtained results show that  the global properties of the 
Simsilun universe asymptotically approach the de Sitter state (cf. Fig. \ref{fig1}). This result {confirms the cosmological no-hair conjecture}, which stipulates that a universe with a non-positive spatial curvature and positive cosmological constant asymptotically approaches the de Sitter state.
On the other hand, the results obtained within the Simsilun simulation, also {confirm the gravitational entropy conjecture} (cf. Fig. \ref{fig2}), which states that the evolution of the universe should be associated with the production on the 
gravitational entropy.

Within the Simsilun simulation, the production of the gravitational entropy is related to the evolution of cosmic structures and presence of virialised objects 
(cf. Fig.~\ref{fig3}). For underdense 
regions the gravitational entropy saturates (i.e. the production rate asymptomatically vanishes, cf. Ref. \cite{2015CQGra..32p5012S} which studied the evolution of cosmic voids and their gravitational entropy). 
The Simsilun Simulation consists of 16,777,216 cells with the average cell's size (at the present-day instant) of a few Mpc. Increasing the resolution and decreasing the size of the cells would require inclusion of several phenomena, which are not included in the Simsilun Simulation but are non-negligible on sub-Mpc scales such as rotation and pressure gradients.

In the Simsilun simulation there is no rotation, nor gradients of pressure which could prevent the collapse \cite{2008MNRAS.391L..59B},
and so the virialisation needs to be 
 externally implemented (cf. \cite{2012JCAP...05..003B,2013JCAP...10..043R,2018A&A...610A..51R}). This is a weak part of the Simsilun simulation and thus the non-zero production rate of the gravitational entropy of the virialised structures should be treated qualitatively. For quantitative results, more realistic simulations are needed, for example the one based on the relativistic Relativistic Zeldovich Approximation (RZA) \cite{2012PhRvD..86b3520B,2013PhRvD..87l3503B,2015PhRvD..92b3512A,2017arXiv171101597R,2015PhRvD..92h3533S,2016JCAP...03..012S}.
The RZA is a general-relativistic approximation that extends the standard
perturbation theory. Recently, it has been shown that the RZA can 
successfully describe collapsing structures and is comparable with Newtonian simulations but includes the relativistic effects \cite{2017AcPPS..10..407O}.

In addition, it should be noted that the Simsilun simulation is based on the Silent Universes which are
Petrov type D. This means that there are no gravitational waves within
the Simsilun simulation. Since recent detections of gravitational radiation 
\cite{2016PhRvL.116f1102A,2016PhRvL.116x1103A,2017PhRvL.118v1101A,2017PhRvL.119p1101A} we 
know  that our Universe should have a large number of sources of gravitational radiation. 
For gravitational waves 
the formula (\ref{CETintgravent}) (the CET case) which was derived for Petrov D does not apply
\cite{2013CQGra..30l5009C}, however formula (\ref{HBMintgravent}) (the HBM case) should still hold.
In addition the gravitational waves deform the spacetime producing the so called {\em memory effect}
\cite{1974SvA....18...17Z,1987Natur.327..123B,1991PhRvL..67.1486C,2016PhRvL.117f1102L,2017PhRvL.118r1103M,2017PhRvD..96f3523K}, which will also contribute to the gravitational entropy. Thus the presence of gravitational waves 
does affect the rate of change of the gravitational entropy.
For example, in the case of Petrov D spacetimes, inside cosmic voids the 
production rate of the gravitational entropy asymptomatically vanishes. Yet, with the 
inclusion of the gravitational waves and the memory effect this may change and lead to a non-zero production rate of the gravitational entropy inside cosmic voids. 
Thus more work is required in the context of the gravitational entropy generated by the gravitational waves. 

In summary, even though the cosmological no-hair and gravitational entropy conjectures appears, at first sight, in contradiction, they 
both correctly  capture properties on a universe with a positive cosmological constant and non-positive spatial curvature.
Therefore, we should expect that our own universe
will keep  producing the gravitational entropy,  even though in the far future its global properties will approach the de Sitter state.

\section*{Acknowledgement}
This work was supported by the Australian Research Council through the Future Fellowship FT140101270. 
The Millennium Simulation databases used in this paper and the web application providing online access to them were constructed as part of the activities of the German Astrophysical Virtual Observatory (GAVO). Computational resources used in this work were provided by the ARC (via FT140101270) and the University of Sydney HPC service (Artemis). 
 Finally, discussions and comments from Thomas Buchert, Timothy Clifton, Alan Coley, Paul Lasky, Jan Ostrowski, Boud Roukema, 
Jos\'e  Senovilla, and Roberto Sussman are gratefully acknowledged.

\bibliography{gravent_rev2} 
\bibliographystyle{ieeetr_arXiv}

\end{document}